\documentclass[fleqn]{article}

\usepackage{amsmath,amssymb}
\usepackage{graphicx}

\def\tlabel#1{\label{#1}}
\def\tref#1#2#3{{#1}~(#2)~#3}

\begin{document}

\title{{\sf  Three Loop Estimate of the Inclusive Semileptonic $b\to c$ Decay Rate}}
\author{M.R.~Ahmady$^1$, F.A.~Chishtie$^2$, V.~Elias$^2$, A.H.~Fariborz$^3$, \\
D.G.C.~McKeon$^2$, T.N.~Sherry$^4$, 
T.G.~Steele$^5$\\[10pt]
\textsl{Theory Group, KEK, Tsukuba, Ibaraki 305-0801, Japan}
}  
\footnotetext[1]{Permanent address: Department of Physics, Mount Allison University, Sackville, NB~~E4L 1E6, Canada}
\footnotetext[2]{Permanent address: Department of Applied Mathematics,
The University of Western Ontario,
London, ON~~ N6A 5B7, Canada}
\footnotetext[3]{Permanent address: Dept.\ of Mathematics/Science, State Univ.\ of New York Institute of Technology, Utica, NY~~13504-3050, USA}
\footnotetext[4]{Permanent address: Department of Mathematical Physics, National University of Ireland, Galway, Ireland }
\footnotetext[5]{Permanent address: Department of Physics \& Engineering Physics,
University of Saskatchewan,
Saskatoon, SK~~ S7N 5E2, Canada}

\maketitle
\begin{abstract}
The renormalization-scale ($\mu$) dependence of the two-loop inclusive semileptonic
$b\to c\ell^-\bar\nu_\ell$ decay rate is shown to be significant in the pole mass scheme, and the decay rate
is shown to be poorly convergent in the $\overline{\rm MS}$ scheme.  Three-loop contributions to the decay 
rate are estimated by developing Pad\'e approximant techniques particularly suited to perturbative calculations 
in the pole mass scheme.   An optimized Pad\'e estimate of the three-loop contributions is obtained by 
comparison of the Pad\'e estimates with the three-loop terms determined by renormalization-group invariance.
The resulting three-loop estimate in the pole-mass scheme 
exhibits  minimal sensitivity to the renormalization scale near
$\mu=1.0\,{\rm GeV}$, leading to an  estimated  decay rate of 
$192\pi^3\Gamma(b\to c\ell^-\bar\nu_\ell)/\left(G_F^2\left|V_{cb}\right|^2\right)=992\pm 217\,{\rm GeV^5}$ inclusive of theoretical  uncertainties and non-perturbative effects.  
\end{abstract}

\section{Introduction}\label{intro_sec}
The CKM matrix element $\left|V_{cb}\right|$, which parametrises one of the sides of the unitarity triangle,
 can be extracted from the inclusive semileptonic decay rate
$\Gamma\left(B\to X_c\ell^-\bar\nu_\ell\right)$.  
From a theoretical perspective, the inclusive process has the 
advantage that non-perturbative contributions are controllable;  hence, an accurate perturbative 
determination of the $b\to c\ell^-\bar\nu_\ell$ decay rate is of value in obtaining $\left| V_{cb}\right|$ 
from data.

Complete two-loop calculations of  semileptonic $b\to c$ decays  exist at the end points 
of the lepton invariant mass  spectrum (maximal and zero recoil) \cite{end_points} and at an intermediate
kinematic value \cite{Czarnecki}.  From these explicit calculations, the total semileptonic decay rate
at two-loop order has been estimated \cite{Czarnecki}.
In this present work, we extend these results to generate an estimate of  the three-loop contributions 
to the  $b\to c\ell^-\bar\nu_\ell$  decay rate via
renormalization-group and Pad\'e-approximant methods.

In Section \ref{two_loop_scale_dep_sec} we demonstrate that the pole mass scheme used in \cite{Czarnecki}
has better perturbative behaviour  than the ${\overline{\rm MS}}$  scheme for the $b\to c$ semileptonic decay rate, a 
result which is somewhat surprising since
the $\overline{\rm MS}$ scheme is better behaved
within calculations of 
the  $b\to u$ semileptonic decay rate
\cite{vanRit}.  The renormalization scale dependence of the  two-loop $b\to c$ rate in the pole mass scheme is extracted using renormalization-group (RG) invariance, and the strong scale dependence that is found serves to motivate 
an estimate of next-order (three-loop) effects.
     
   The scale dependence of the decay rate is an important component of the
procedure for estimating three-loop corrections to the $b\to  c$ semileptonic
rate. Using information obtained from RG invariance, Pad\'e approximation
methods are developed in Section \ref{pade_sec}  that are appropriate for estimating
next-order terms within pole-mass perturbative calculations.  An optimized
Pad\'e estimate of the three-loop constant coefficient ({\it i.e.} the
non-logarithmic term) is obtained by finding the best agreement between
Pad\'e estimates and true values of the RG-accessible three-loop coefficients
of logarithms. The RG-determinations of these latter coefficients in
conjunction with the optimized Pad\'e estimate of the constant coefficient
together constitute a scale-sensitive estimate of the full three-loop
contribution to the perturbative rate.

This Pad\'e estimate of aggregate three-loop effects allows
the renormalization-scale dependence of the $b\to c\ell^-\bar\nu_\ell$ decay rate to be studied.  
In Section \ref{three_loop_scale_dep_sec}, 
a region of minimal scale sensitivity  \cite{PMS} is found in the resulting decay  rate.  This 
minimal-sensitivity scale  is found to be
close to  the fastest-apparent-convergence renormalization scale at which the three-loop contributions 
vanish entirely  \cite{Grunberg}.   The proximity of these two scales 
supports the validity of these scales for obtaining
estimates of the three-loop perturbative  $b\to c\ell^-\bar\nu_\ell$ decay rate.  Theoretical uncertainties 
in this estimate are considered in Section \ref{discuss_sec}.

\section{Scale Dependence of the Two-Loop Rate}\label{two_loop_scale_dep_sec}
In ref.\ \cite{Czarnecki}, the inclusive semileptonic $b\to c\ell^-\bar\nu_\ell$ rate 
is estimated to two-loop order in
terms of renormalization-group- (RG-) invariant pole masses $m_b$ and $m_c$. The rate $\Gamma$ can be
expressed in the form
\begin{equation}
\Gamma=\frac{G_F^2m_b^5\left|V_{cb}\right|^2}{192\pi^3}F\left(\frac{m_c^2}{m_b^2}\right)
S\left[\frac{\alpha_s(\mu)}{\pi},\log{\left(\frac{\mu^2}{m_bm_c}\right)}\right]
\tlabel{rate}
\end{equation}
with the RG-invariant form-factor
\begin{equation}
F(r)=1-8r-12r^2\log(r)+8r^3-r^4
\tlabel{form_fac}
\end{equation}
preceding the perturbative series $S$ whose two-loop contribution at $\mu^2=m_bm_c$    
has been estimated in \cite{Czarnecki} by combining explicit results at the end  and intermediate points 
\cite{end_points,Czarnecki}
of the 
decay spectrum:\footnote{The ${\cal O}\left(\alpha_s^2\right)$ coefficient quoted in 
(\protect\ref{two_loop_rate}) [A. Czarnecki, personal communication] has been slightly corrected from the value $-8.4\pm 0.4$ appearing in ref.\ \protect\cite{Czarnecki}.}\label{czarnecki_note}
\begin{equation}
S\left[\frac{\alpha_s\left(\sqrt{m_bm_c}\right)}{\pi},0\right]=1
-1.67\frac{\alpha_s\left(\sqrt{m_bm_c}\right)}{\pi}
-\left(8.9\pm 0.3\right)\left[\frac{\alpha_s\left(\sqrt{m_bm_c}\right)}{\pi}\right]^2~~.
\tlabel{two_loop_rate}
\end{equation}
If we assume central values $\alpha_s\left(m_\tau\right)= 0.33$ \cite{aleph,steele} 
and $m_b = 4.9\, {\rm GeV}$ \cite{hoang}, and if we follow ref.\ \cite{Czarnecki}
in assuming that $m_c = 0.3 m_b$, we find that  $\alpha_s$ evolves from $\mu=m_\tau$ 
through the four-loop, four-flavour
\footnote{The use of four contributing flavours is necessary since  $\alpha_s$ is referenced to four flavours in (\protect\ref{two_loop_rate}) \protect\cite{Czarnecki}.}
 ($n_f=4$) $\beta$-function \cite{beta}
to $  \alpha_s\left(\sqrt{m_bm_c}\right)/\pi=0.087$,
in which case
$S\left[0.087,0\right]=1-0.145-\left(0.067\pm 0.002\right)$.
  It is evident from (\ref{two_loop_rate}) that truncation after two-loop order introduces theoretical
uncertainty of order $8.5\% (= 0.067/0.79)$. 

There are, of course, other sources of theoretical
uncertainty.  Ambiguities concerning the definition of a pole mass in the presence of
nonperturbative (confinement) effects have led the authors of ref.\ \cite{Czarnecki} to reparametrise the rate
(\ref{rate}) in terms of ``low-scale'' masses obtained through additional phenomenology--- we will
ultimately compare our results to the reparametrised rate in Section \ref{discuss_sec}. In light of more recent
work \cite{vanRit,hoang,pole_mass1,pole_mass2}  
specifying accurate pole mass values by relating pole masses to $\overline{\rm MS}$ and  $\Upsilon$-scheme
masses with three-loop precision, we take a more empirical approach to the utilisation of pole
masses within inclusive semileptonic rates.  

   Renormalization-scale dependence provides an additional source of theoretical
uncertainty to any rate calculated via the series (\ref{two_loop_rate}). If $S$ 
 varies for different choices
of $\mu$, a value for $\Gamma$  extracted from any particular choice of $\mu$ ({\it e.g.,} $\mu^2 =m_bm_c$) 
is compromised. 
The optimal value of $\mu$ has been argued to be the choice for which $S$ has minimal
sensitivity to the renormalization scale \cite{PMS}
{\it i.e.,} the point at which $\mathrm{d}S/\mathrm{d}\mu = 0$. For (\ref{two_loop_rate}) to lead to 
a reliable estimate of the true semileptonic $b\to c$ rate, one would not necessarily need to establish
that $\mu^2=m_bm_c$  is such a ``Principle of Minimal Sensitivity'' (PMS) point, but rather that the rate
calculated at this value of $\mu$ differs only inconsequentially from the rate calculated at the true PMS
value of $\mu$. Any discrepancy between rates calculated at  $\mu^2=m_bm_c$   and at  $\mu_{PMS}$  is a 
direct measure of the former rate's theoretical uncertainty arising from renormalization-scale ambiguities.

    The two-loop order renormalization-scale dependence implicit in the perturbative series
within (\ref{rate}) may be parametrised as follows:
\begin{gather}
S\left[x(\mu),L(\mu)\right]=1+\left(a_0+a_1L\right)x+\left(b_0+b_1L+b_2L^2\right)x^2~,
\tlabel{S_def}\\
a_0=-1.67~,~b_0=-8.9\pm 0.3~,~x(\mu)\equiv\frac{\alpha_s(\mu)}{\pi}
~,~L(\mu)\equiv\log{\left(\frac{\mu^2}{m_bm_c}\right)}~.
\tlabel{misc_defs}
\end{gather}
The renormalization-scale invariance of the all orders rate implies that
\begin{equation}
\mu^2\frac{\mathrm{d}\Gamma}{\mathrm{d}\mu^2}=0
\tlabel{RG_inv}
\end{equation}
hence that,
\begin{equation}
0=\frac{\partial S\left[x,L\right]}{\partial L}-\left(\beta_0x^2+\beta_1 x^3+\beta_2 x^4+\ldots\right)
\frac{\partial S\left[x,L\right]}{\partial x}
\tlabel{RG_eqn}
\end{equation}
where the normalization of QCD  $\beta$-function coefficients is explicitly defined by
\begin{equation}
\mu^2\frac{\mathrm{d}x}{\mathrm{d}\mu^2}=-\left(\beta_0x^2+\beta_1 x^3+\beta_2 x^4+\ldots\right)
\end{equation}
with $n_f = 4$ values  $\beta_0 = 25/12$,  $\beta_1 = 77/24$, and  $\beta_2 = 21943/3456$ 
(in the  $\overline{\rm MS}$ scheme). It is easily seen 
from (\ref{RG_eqn}) that the logarithmic coefficients within (\ref{S_def}) are
\begin{equation}
a_1=b_2=0~,~b_1=a_0\beta_0= -3.479~.
\tlabel{RG_2loop}
\end{equation}
In Figure \ref{vcb_2loop_fig}, we have plotted the mass- and scale-dependent portion of the rate (\ref{rate}),
\begin{equation}
\frac{\Gamma}{K}=m_b^5\,F\left(\frac{m_c^2}{m_b^2}\right)\,S\left[x(\mu),L(\mu)\right]\quad ,\quad
\left(K\equiv \frac{192\pi^3}{G_F^2\left|V_{cb}\right|^2}\right)~,
\tlabel{red_rate_def}
\end{equation}
as a function of $\mu$. 
The Fig.\ \ref{vcb_2loop_fig} curve is obtained by assuming $m_b=4.9\,{\rm GeV}$ \cite{hoang},
$m_c=0.3m_b$ \cite{Czarnecki}, and the central value $b_0=-8.9$ from the estimate \cite{Czarnecki} given in 
(\ref{misc_defs}). 
Substantial scale dependence is evident from the figure: the rate increases
monotonically with $\mu$, flattening out somewhat for larger values.  Moreover the curve exhibits no
PMS point ({\it i.e.,} extremum), and at $\mu =\sqrt{m_cm_b}= 2.68 \,{\rm GeV}$ 
yields a rate  10\% smaller than the (still
increasing) rate at $\mu = 8 \,{\rm GeV}$, the flattest portion of the curve shown in the figure. 
Thus, the two-loop pole-mass calculation of the  $b\to c\ell^-\bar\nu_\ell$ rate exhibits at 
best  only poorly-controllable
dependence on the choice of renormalization scale. Such scale dependence may
compromise any subsequent  ``low-scale'' mass expression devolving from the two-loop 
$\mu =\sqrt{m_bm_c}$    pole-mass rate. 
     
Similar scale dependence and even worse apparent non-convergence characterise the 
two-loop order pole-mass calculation of the $b\to u\ell^-\bar\nu_\ell$ rate, thereby motivating a recasting of the
calculation in terms of the $\overline{{\rm MS}}$ running $b$-quark mass \cite{vanRit}. 
For four contributing flavours, 
the two-loop relationship between pole and  $\overline{{\rm MS}}$ 
running quark masses is \cite{pole_mass1} 
\begin{equation}
m^{pole}=m(\mu)\left[
1+\left(\frac{4}{3}+\log{\left(\frac{\mu^2}{m^2(\mu)}\right)}\right)x(\mu)
+\left(10.3919+\frac{415}{72}\log{\left(\frac{\mu^2}{m^2(\mu)}\right)}
+\frac{37}{24}\log^2{\left(\frac{\mu^2}{m^2(\mu)}\right)}
\right)x^2(\mu)
\right]
\tlabel{pole_mass}
\end{equation}
By substituting this relation into (\ref{rate}), (\ref{S_def}), and (\ref{misc_defs}), we find that the fully 
$\overline{{\rm MS}}$ version of the
mass- and scale-dependent portion (\ref{red_rate_def}) of the $b\to c\ell^-\bar\nu_\ell$ 
decay rate (using the central value $b_0=-8.9$) is
\begin{equation}
\begin{split}
\frac{\Gamma^{\overline{MS}}}{K}=m_b^5(\mu)&\,F\left(\frac{m_c^2}{m_b^2}\right)
\Biggl\{
1+\left(5.00+5\log\left(\frac{\mu^2}{m_b^2(\mu)}\right) \right)x(\mu)
\\
&+\left[49.3-1.74\,\log\left(\frac{\mu^2}{m_c^2(\mu)}\right)+45.4\,\log\left(\frac{\mu^2}{m_b^2(\mu)}\right)
+\frac{425}{24}\,\log^2\left(\frac{\mu^2}{m_b^2(\mu)}\right)
  \right]x^2(\mu)
\Biggr\}~.
\end{split}
\tlabel{MS_rate}
\end{equation}
Note that $F\left(m_c^2/m_b^2\right)$ remains RG invariant if
$m_c\to m_c(\mu)$ and $m_b\to m_b(\mu)$. If $\mu^2=m_b(\mu)m_c(\mu)$
and if we continue to assume that $m_c(\mu)=0.3m_b(\mu)$, then
\begin{equation}
\frac{\Gamma^{\overline{MS}}}{K}=m_b^5(\mu)
\,F\left(0.09\right)
\left[
1-1.02\, x\left(\mu\right)+18.2\, x^2\left(\mu\right)
\right]\quad ,\quad \mu=\sqrt{0.3}m_b(\mu)
\tlabel{MS_rate_value}
\end{equation}
The convergence of this $\overline{{\rm MS}}$ perturbative series is even more ill-behaved 
than its pole-mass
version (\ref{two_loop_rate}).  Moreover, the scale dependence of (\ref{MS_rate}) is shown in 
Figure \ref{vcb_ms_2loop_fig} to be even more
pronounced than that of the same rate in the pole-mass scheme 
(Figure \ref{vcb_2loop_fig})--- Figure \ref{vcb_ms_2loop_fig} displays a
rate which decreases with $\mu$ with no apparent PMS point. 

     Thus the $\overline{{\rm MS}}$ approach, which substantially improves the perturbative series within the
semileptonic $b\to   u$ rate \cite{vanRit}, fails to improve the pole-mass expressions (\ref{rate}--\ref{two_loop_rate}) for the
semileptonic $b\to   c$ rate.  If this latter rate is to be utilised to extract an estimate of
$\left|V_{cb}\right|$ from
the inclusive  $B\to X_c\ell^-\bar\nu_\ell$   branching ratio, there is evident value in
having an estimate of next-order corrections  in order to obtain some control over
renormalization-scale dependence. This three-loop order contribution to $S\left[x(\mu),L(\mu)\right]$ is 
necessarily of the
form
\begin{equation}
\Delta S^{3L}\left[x(\mu),L(\mu)\right]=\left[
c_0+c_1L(\mu)+c_2L^2(\mu)+c_3L^3(\mu)
\right]x^3(\mu)
\tlabel{three_loop_term}
\end{equation}
The RG equation (\ref{RG_eqn}) implies that
\begin{equation}
\begin{split}
0=&a_1x+\left(b_1-a_0\beta_0\right)x^2+\left(2b_2-\beta_0a_1\right)x^2L+
\left[c_1-2\beta_0b_0-a_0\beta_1\right]x^3
+\left[2c_2-2\beta_0b_1-\beta_1a_1\right]x^3L
\\
&+
\left[3c_3-2\beta_0b_2\right]x^3L^2+{\cal O}\left(x^4\right)~.
\end{split}
\tlabel{RG_three_loop}
\end{equation}
The set of results (\ref{RG_2loop}) are evident from the requirement that ${\cal O}(x)$, 
${\cal O}\left(x^2\right)$ and ${\cal O}\left(x^2L\right)$ terms in (\ref{RG_three_loop})
separately vanish. The vanishing of subsequent terms in (\ref{RG_three_loop}) implies that 
\begin{equation}
c_1=2b_0\beta_0+a_0\beta_1~,~c_2=a_0\beta_0^2~,~c_3=0~.
\tlabel{an_RG_coeffs}
\end{equation}
For $a_0$ and $b_0$ as given in (\ref{misc_defs}) and 
three massless flavours (as appropriate to phenomenology devolving 
from $\alpha_s\left(m_\tau\right)$ \cite{aleph}),
\begin{equation}
c_1=-42.4\pm 1.3~,~c_2= -7.25~,~c_3=0~.
\tlabel{RG_coeffs}
\end{equation}
The coefficient $c_0$, however, is RG-inaccessible to these orders of perturbation theory, and
requires a direct three-loop calculation. In the absence of such a calculation, we estimate $c_0$ in the
section which follows via asymptotic Pad\'e approximant methodology, in much the same way as
in a prior estimate \cite{V_ub} of the three-loop contribution to the  
$b\to u\ell^-\bar\nu_\ell$ decay rate.

\section{RG/Pad\'e Estimate of $c_0$}\label{pade_sec}
Consider a perturbative field-theoretical series with $N+M$ known terms:
\begin{equation}
S(x)=1+R_1x+R_2x^2+\ldots+R_{N+M}x^{N+M} +\ldots~~.
\tlabel{basic_series}
\end{equation}
The set of known coefficients $\{R_1,R_2,\ldots ,R_{N+M}\}$ is sufficient to determine in full the $N+M$
coefficients characterising an $[N|M]$ Pad\'e-approximant to the series $S$:
\begin{equation}
S_{[N|M]}(x)\equiv \frac{1+a_1x +a_2 x^2+\ldots+a_Nx^N}{1+b_1x +b_2 x^2+\ldots+b_Mx^M}~.
\tlabel{basic_pade}
\end{equation}
The coefficients $\{a_1,\ldots ,a_N,b_1,\ldots , b_M\}$ are obtained by the requirement that the 
power-series expansion of $S_{[N|M]}(x)$ recovers the $N+M$ known coefficients within (\ref{basic_series}), the series  $S(x)$. The next ${\cal O}\left(x^{N+M+1}\right)$ term in this power series is a Pad\'e approximant 
prediction for the first  unknown coefficient $R_{N+M+1}$.  For example, if only the next-to-leading order 
coefficient $R_1$ is known, one can use this coefficient to construct a $[0|1]$ approximant to the series
(\ref{basic_series}),
\begin{equation}
S_{[0|1]}(x)=\frac{1}{1-R_1x}=1+R_1x+R_1^2x^2+\ldots~,
\tlabel{S_01}
\end{equation}
for which $R_1^2$ is the {\em predicted} value of $R_2$:
\begin{equation}
R_2^{[0|1]}=R_1^2~.
\tlabel{R2_01}
\end{equation}
Similarly, if $R_1$ and $R_2$ are known (corresponding to two subleading orders of perturbation theory),
one has enough information to construct a $[1|1]$ approximant to the series (\ref{basic_series}):
\begin{equation}
S_{[1|1]}(x)=\frac{1+\left(\frac{R_1^2-R_2}{R_1}\right)x}{1-\frac{R_2}{R_1}x}
=1+R_1x+R_2x^2+\frac{R_2^2}{R_1}x^3+\ldots~ .
\tlabel{S_11}
\end{equation}
The predicted value for the unknown coefficient $R_3$ is 
\begin{equation}
R_3^{[1|1]}=R_2^2/R_1
\tlabel{R3_11}
\end{equation}

In general, one can always use an $[N|M]$ approximant (\ref{basic_pade}) to predict the first unknown series coefficient $R_{N+M+1}$.  
Such predictions have accuracy which increases as $N$ and $M$ increase.  For perturbative field-theoretical series (characterised by asymptotic  $R_N\sim N!$ behaviour) the accuracy of such predictions has been  argued by Ellis, Karliner and Samuel to satisfy the relative error formula  \cite{Samuel}
\begin{equation}
\frac{\Delta R_{N+M+1}^{[N|M]}}{R^{true}_{N+M+1}}=\frac{R_{N+M+1}^{[N|M]}-R_{N+M+1}^{true}}{R^{true}_{N+M+1}}
=-\frac{M!A^M}{\left(N+M+aM+b\right)^M}
\tlabel{err_form}
\end{equation}
where $A$, $a$, and $b$ are constants to be determined.
Of particular interest are the relative errors obtained from (\ref{err_form}) for the predictions (\ref{R2_01})
and (\ref{R3_11})
\begin{gather}
\frac{R_1^2-R_2}{R_2}=-\frac{A}{1+(a+b)}~,
\tlabel{delta_R2}\\
\frac{\frac{R_2^2}{R_1}-R_3}{R_3}=-\frac{A}{2+(a+b)}~.
\tlabel{delta_R3}
\end{gather}
Denoting the constant $a+b=k$, we can eliminate the other constant $A$ within (\ref{delta_R2},\ref{delta_R3}) 
and solve for $R_3$ algebraically 
to obtain the improved estimate
\begin{equation}
R_3=\frac{(2+k)R_2^3}{(1+k)R_1^3+R_1R_2}~.
\tlabel{R3_k}
\end{equation}
The assumption $k=0$ has been utilised in prior applications to predict successfully the third subleading 
contribution to the QCD $\overline{{\rm MS}}$ $\beta$-function \cite{Samuel}, the DRED SQCD $\beta$-function 
\cite{DRED}, the  $\overline{{\rm MS}}$ $\beta$-function for massive scalar field theory \cite{Chishtie}, as well as to obtain estimates of such contributions for a number of processes calculated in the  $\overline{{\rm MS}}$
scheme: SM \cite{SM_higgs} and MSSM \cite{MSSM_higgs} Higgs $\to$ 2 gluon decay rates, 
$b\to u \ell^-\bar\nu_\ell$ \cite{V_ub}, $W^+W^-\to ZZ$ \cite{Chishtie}, 
and the QCD static potential function \cite{static}.

The choice $k=0$, however, is ill-suited to pole-mass calculations.  To see this, consider the series 
(\ref{S_def}, \ref{three_loop_term}) within 
(\ref{rate}) characterising the pole-mass expression for the $b\to c\ell^-\bar \nu_\ell$ decay rate
\begin{gather}
S[x,L]=1+\left(a_0+a_1L\right)x+\left(b_0+b_1L+b_2 L^2\right)x^2
+\left(c_0+c_1L+c_2L^2+c_3L^3\right)x^3
\tlabel{pert_form}\\
x=\frac{\alpha_s(\mu)}{\pi}\quad ,\quad L=\log\left(\frac{\mu^2}{m_bm_c}\right)
\nonumber
\end{gather}
We have already seen that $a_0=-1.67$, $a_1=0$, $b_0\cong -8.9\pm 0.3$, $b_1=a_0\beta_0= -3.48$, $b_2=0$.  The 
above series is in the form of (\ref{basic_series}) with $R_1$ and $R_2$  respectively identified with 
$a_0+a_1L$ and $b_0+b_1L+b_2L^2$.  If $k=0$, we see from (\ref{R3_k}) that 
\begin{equation}
R_3=\frac{2\left(b_0+b_1L\right)^3}{a_0\left(a_0^2+b_0+b_1L\right)}=\frac{2b_1^2}{a_0}L^2+
\frac{2b_1}{a_0}\left(2b_0-a_0^2\right)L+{\cal O}\left(L^0\right)
\tlabel{R3_k0}
\end{equation} 
Comparing this expression to the form (\ref{three_loop_term}) anticipated for the third subleading order of $S[x,L]$, 
we necessarily obtain  the following predictions for the RG accessible coefficients $c_2$, $c_1$:
\begin{gather}
c_2=2\frac{b_1^2}{a_0}=2a_0\beta_0^2~,
\tlabel{c2_k0}\\
c_1=2\left(2b_0-a_0^2\right)\frac{b_1}{a_0}=4\beta_0b_0-2\beta_0a_0^2
\tlabel{c1_k0}
\end{gather}
It is evident from (\ref{an_RG_coeffs},\ref{RG_coeffs}) 
that these predictions are quite poor; (\ref{c2_k0}) is double the true 
value for $c_2$, as obtained in (\ref{RG_coeffs}), and $c_1$ is also badly overestimated
[for values (\ref{RG_2loop}) and the central value $b_0=-8.9$, eq.\ (\ref{c1_k0}) implies that $c_1=-85.8$, in contrast to the correct (RG) 
value  $c_1=-42.4$].  

For a given choice of $k$, estimates of the  coefficients $c_i$ characterising the third subleading order have 
been obtained for correlation functions \cite{corr} and the $b\to u \ell^-\bar\nu_\ell$ rate \cite{V_ub} by 
moments of $R_3$, as estimated in (\ref{R3_k}), over the entire ultraviolet region
({\it e.g.,}  $\mu^2/(m_b m_c)\ge 1$ for the case at hand). These moments are then equated to corresponding moments of
$R_3=c_0+c_1L+c_2L^2+c_3L^3$ in order to obtain values for $\{c_0,c_1,c_2,c_3\}$.  If we define $w=m_bm_c/\mu^2$ ($L=-\log w$), the moments 
\begin{equation}
N_j=(j+2)\int\limits_0^1w^{j+1}R_3(w)\mathrm{d}w
\tlabel{moments}
\end{equation}
can be obtained using (\ref{R3_k}) for the integrand $R_3$, which becomes a function of $w$ for our pole 
mass case by virtue of the $w$-dependence of $R_2$: $R_1=a_0$, $R_2=b_0-b_1\log w$.  After (numerical) 
computation of the 
values of $N_j$,  the coefficients $c_i$ are  obtained 
by equating such values to
 the corresponding integrals
\begin{equation}
N_j=(j+2)\int\limits_0^1w^{j+1}\left(c_0-c_1\log w+c_2\log^2w-c_3\log^3w\right)\mathrm{d}w~.
\tlabel{moment_eqns}
\end{equation}
In particular, we see that
\begin{gather}
N_{-1}=c_0+c_1+2c_2+6c_3
\tlabel{Nm1}\\
N_0=c_0+\frac{1}{2}c_1+\frac{1}{2}c_2+\frac{3}{4}c_3
\tlabel{N0}\\
N_1=c_0+\frac{1}{3}c_1+\frac{2}{9}c_2+\frac{2}{9}c_3
\tlabel{N1}\\
N_2=c_0+\frac{1}{4}c_1+\frac{1}{8}c_2+\frac{3}{32}c_3~.
\tlabel{N2}
\end{gather}

If $k\ne -1$, such a procedure is seen to lead to a non-zero value of $c_3$, in contradiction to the result 
$c_3=0$ (\ref{an_RG_coeffs}) necessarily following from application of the RG equation (\ref{RG_eqn}) within the pole mass scheme.  
For the case $k=-1$, however, the result (\ref{R3_k}) collapses to the naive estimate (\ref{R3_11}), which in the pole mass scheme is necessarily a degree-2 polynomial in $L$:
\begin{equation}
R_3^{(k=-1)}=\frac{R_2^2}{R_1}=\frac{1}{a_0}\left(b_0+b_1L\right)^2=
\frac{b_1^2}{a_0}L^2+2\frac{b_1b_0}{a_0}L+\frac{b_0^2}{a_0}
\tlabel{R3_km1}
\end{equation}
The moment procedure  described above then reduces to fitting the degree-3 polynomial 
$R_3=c_3L^3+c_2L^2+c_1L+c_0$ to the degree-2 polynomial in (\ref{R3_km1}).  Such a fit necessarily reduces to  
 equating the powers of $L$ in these two expressions.  We thus find that $c_3$ must be  zero, and that
\begin{gather}
c_2=\frac{b_1^2}{a_0}=a_0\beta_0^2
\tlabel{c2_km1}\\
c_1=2\frac{b_1b_0}{a_0}=2\beta_0b_0
\tlabel{c1_km1}\\
c_0=\frac{b_0^2}{a_0}
\tlabel{c0_km1}
\end{gather}
Equation (\ref{c2_km1}) is in exact agreement with the result (\ref{an_RG_coeffs}) 
obtained via RG-invariance from 
(\ref{RG_eqn}).  Equation (\ref{c1_km1}) represents the first (and dominant) contribution to the RG-determination 
of $c_1$ in (\ref{an_RG_coeffs}).  For the central value $b_0=-8.9$ (and $n_f=4$) the (\ref{c1_km1}) prediction 
$c_1=-37.1$ is not far from the true value $c_1=-42.4$.  The accuracy of these results provide some support to the 
naive estimate $c_0=-47.4$ 
obtained via (\ref{c0_km1}) for the RG-inaccessible coefficient $c_0$.

We can further improve our estimate for $c_0$ by finding the value of $k$ within (\ref{R3_k}) which, when used 
within the integrand of (\ref{moments}) to match the moments $N_j$ to (\ref{Nm1}--\ref{N2}), most closely 
reproduces
the true values of $c_2$ and $c_1$ , as determined by RG methods in (\ref{an_RG_coeffs}) and (\ref{RG_coeffs}). 
In such
a procedure we obtain estimated values for $c_3$, $c_2$, $c_1$, and $c_0$ that depend explicitly on $k$:
\begin{gather}
c_0(k)=-\frac{1}{6}N_{-1}(k)+4N_0(k)-\frac{27}{2}N_1(k)+\frac{32}{3}N_2(k)
\tlabel{c0_k}\\
c_1(k)=\frac{3}{2}N_{-1}(k)-32N_0(k)+\frac{189}{2}N_1(k)-64N_2(k)
\tlabel{c1_k}\\
c_2(k)=-\frac{13}{6}N_{-1}(k)+38N_0(k)-\frac{189}{2}N_1(k)+\frac{176}{3}N_2(k)
\tlabel{c2_k}\\
c_3(k)=\frac{2}{3}N_{-1}(k)-8N_0(k)+18N_1(k)-\frac{32}{3}N_2(k)
\tlabel{c3_k}
\end{gather}
where
\begin{equation}
N_j(k)=(j+2)\int\limits_0^1w^{j+1}
\left[\frac{(2+k)\left(b_0-b_1\log w\right)^3}{(1+k)a_0^3+a_0\left(b_0-b_1\log w\right)}\right]\,\mathrm{d}w
\tlabel{Nj_k}
\end{equation}
We then use the explicit values for $c_1$ and $c_2$ obtained from (\ref{RG_coeffs})  by RG methods to 
optimize the sum of the squares of the relative error of  estimated values of $c_1$ and $c_2$,
\begin{equation}
\Delta(k)=\left(\frac{c_2(k)-a_0\beta_0^2}{a_0\beta_0^2}\right)^2+
\left(\frac{c_1(k)-\left(2\beta_0b_0-a_0\beta_1\right)}{2\beta_0b_0-a_0\beta_1}\right)^2
\tlabel{Delta}
\end{equation}
with respect to $k$.  For the set of values $a_0=-1.67$, $b_0=-8.9$, $\beta_0=25/12$, $\beta_1=77/24$, we find a 
clear minimum of $\Delta(k)$ at $k=-0.94$, as evident from Figure \ref{vcb_k_opt_fig}, consistent with 
the case made in the preceding paragraph for 
 an optimal $k$ value  close to $k=-1$. Corresponding $k=-0.94$ values for 
the moments $N_j$ are 
obtained numerically via (\ref{Nj_k})
\begin{equation}
N_{-1}=-106.3~,~N_0=-74.92~,~ N_1=-66.17~,~N_2=-62.12
\tlabel{moment_values}
\end{equation}
and these values lead via (\ref{c0_k})--(\ref{c3_k})to the following estimates for the third subleading order coefficients:
\begin{equation}
c_3=2.0\times 10^{-4}~,~c_2=-7.68~,~c_1=-39.7~,~c_0=-51.2
\tlabel{coeff_values}
\end{equation}
These  values  reflect excellent agreement with the RG values (\ref{RG_coeffs}).
The above estimate for the RG-inaccessible coefficient  $c_0$ is only 20\% larger 
in magnitude than the naive prediction (\ref{c0_km1}), indicative of the internal consistency of the methodology.

We conclude by noting that four separate estimates of $c_0$ can be obtained from (\ref{Nm1})--(\ref{N2}) by 
substituting into these equations the optimal $N_j$ values (\ref{moment_values}) as well as the true values
$c_3=0$, $c_1=-42.4$ and $c_2=-7.25$ as determined by RG invariance (\ref{RG_coeffs}) in the previous section.  
Solving each equation separately for 
$c_0$, we find that
\begin{gather}
c_0=N_{-1}-c_1-2c_2-6c_3=-49.4~,
\tlabel{c0_opt_m1}\\
c_0=N_0-\frac{1}{2}c_1-\frac{1}{2}c_2-\frac{3}{4} c_3=-50.1~,
\tlabel{c0_opt_0}\\
c_0=N_1-\frac{1}{3}c_1-\frac{2}{9}c_2-\frac{2}{9} c_3=-50.4~,
\tlabel{c0_opt_1}\\
c_0=N_2-\frac{1}{4}c_1-\frac{1}{8}c_2-\frac{3}{32} c_3=-50.6~.
\tlabel{c0_opt_2}
\end{gather}
The estimates (\ref{c0_opt_m1})--(\ref{c0_opt_2}) are remarkably consistent with each other and only slightly 
smaller in magnitude than the estimate in (\ref{coeff_values}) 
obtained without RG inputs for $\{c_1,~c_2~,c_3\}$.

In estimating the third subleading order of 
(\ref{three_loop_term}), the series within the rate (\ref{rate}), 
 we choose (\ref{c0_opt_0}) as the most 
central  of our $c_0$ estimates, in conjunction with the explicit RG determinations of $c_1$ and $c_2$ ($c_3=0$).  
This set of values is obtained, as noted earlier, for the central value estimate \cite{Czarnecki} of $b_0$:
\begin{equation}
b_0=-8.9~:~\{c_0,c_1,c_2\}\cong\{-50.1,-42.4,-7.25\}~.
\tlabel{central_coeffs}
\end{equation}
Precisely the same procedure can be used to obtain corresponding estimates of $\{c_0,c_1,c_2\}$ for the extremes of the $b_0=-8.9\pm0.3$ range obtained in \cite{Czarnecki} (see footnote \ref{czarnecki_note}):
\begin{gather}
b_0=-9.2~:~\{c_0,c_1,c_2\}\cong\{-53.6,-43.7,-7.25\}
\tlabel{high_coeffs}\\
b_0=-8.6~:~\{c_0,c_1,c_2\}\cong\{-47.5,-41.2,-7.25\}~.
\tlabel{low_coeffs}
\end{gather}
These estimates are obtained in precisely the same way as those of (\ref{central_coeffs}): $c_1$ and $c_2$ are 
identified with their RG values via (\ref{RG_coeffs}), and $c_0$ is determined via (\ref{c0_opt_0}) with $N_0(k)$ [see 
Eq.\ (\ref{Nj_k})] evaluated at a value of $k$ which minimizes $\Delta(k)$ [see Eq.\ (\ref{Delta})].  
The coefficient $c_3=0$ for all cases, as evident in the previous section from RG-invariance.    
For the $b_0=-9.2$ case, the minimizing value  $k=-0.94$ is the same as for $b_0=-8.9$.  For $b_0=-8.6$, the minimum value of $\Delta(k)$ 
is found to occur when  $k=-0.93$.

\section{Scale Dependence of the Three-Loop Rate}\label{three_loop_scale_dep_sec}
\begin{figure}[hbt]
\centering
\includegraphics[scale=0.75]{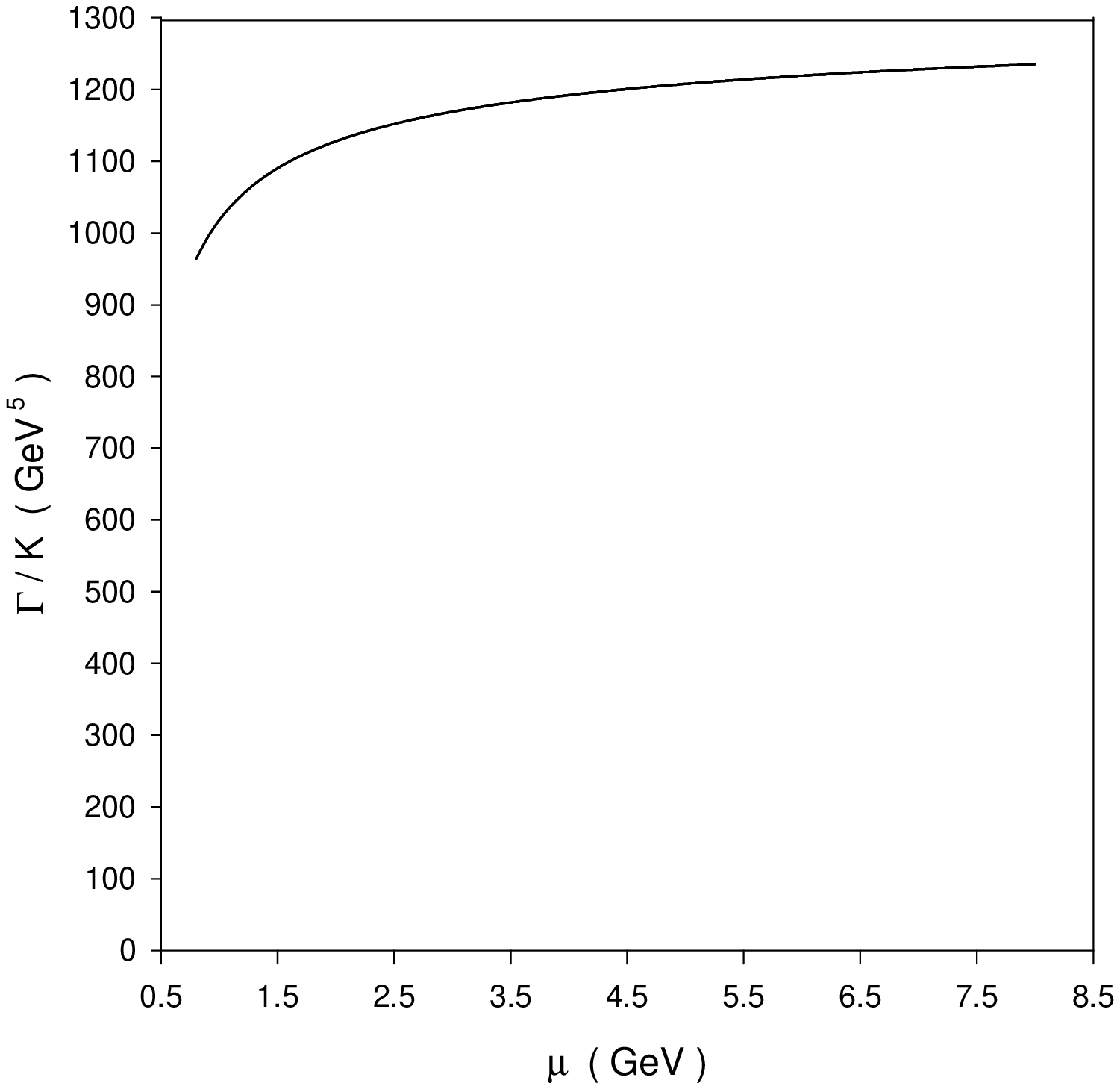}
\caption{Renormalization scale ($\mu$) dependence of the two-loop reduced rate $\Gamma/K$ in the pole mass scheme.
}
\label{vcb_2loop_fig}
\end{figure}

\begin{figure}[hbt]
\centering
\includegraphics[scale=0.75]{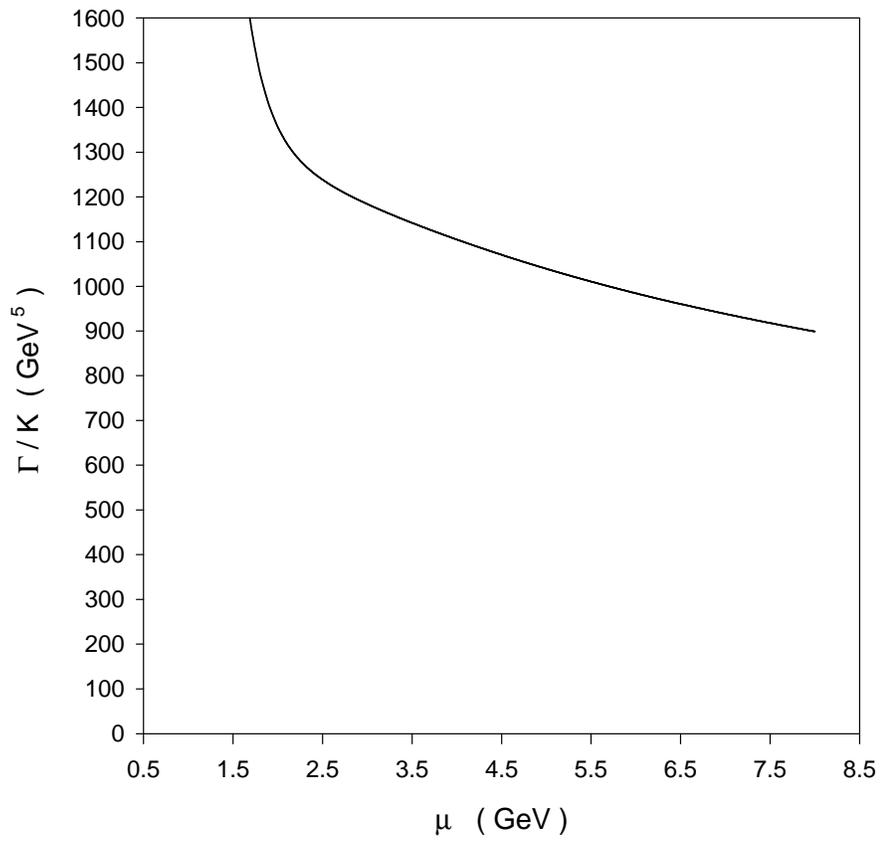}
\caption{Renormalization scale ($\mu$) dependence of the two-loop reduced rate  $\Gamma/K$ in the $\overline{\rm MS}$ scheme.  In the $\overline{\rm MS}$ scheme,  
$m_b(\mu)$ is obtained from the four-loop $n_f=4$ anomalous mass dimension 
\protect\cite{chetyrkin}  
using $m_b\left(m_b\right)=4.2\,{\rm GeV}$ \cite{hoang} as a reference value. 
}
\label{vcb_ms_2loop_fig}
\end{figure}

\begin{figure}[hbt]
\centering
\includegraphics[scale=0.75]{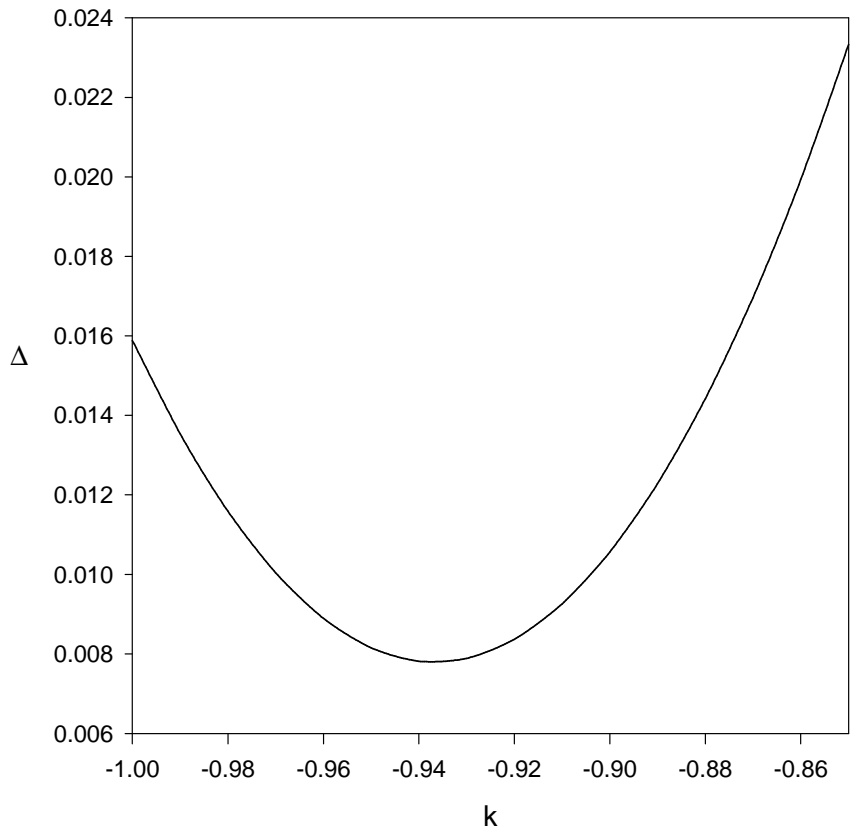}
\caption{The quantity $\Delta$ measuring the sum of the  relative errors in the Pad\'e 
estimate 
(\protect\ref{c1_k},\protect\ref{c2_k})
of the RG-accessible three-loop coefficients $\{c_1,c_2\}$  plotted as a function of the 
error-formula parameter $k$ (\protect\ref{R3_k}). The location of the curve's minimum represents the  value of $k$ 
leading to an optimized Pad\'e estimate. 
}
\label{vcb_k_opt_fig}
\end{figure}

\begin{figure}[hbt]
\centering
\includegraphics[scale=0.75]{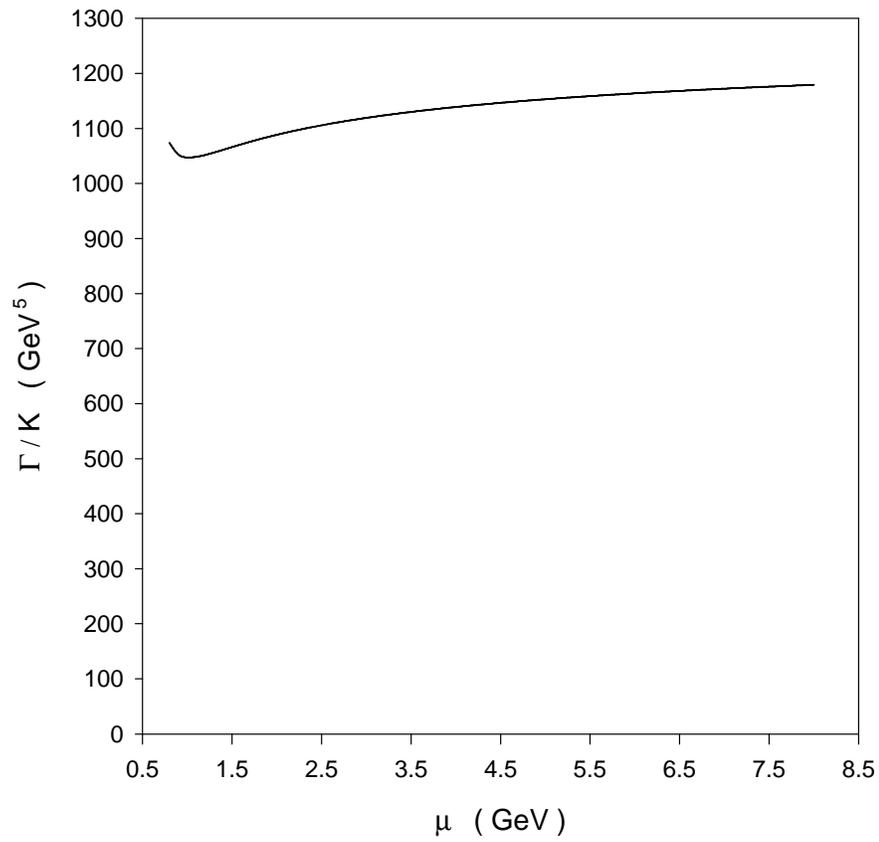}
\caption{Renormalization scale ($\mu$) dependence of the three-loop reduced rate  $\Gamma/K$ 
in the pole mass scheme.   The PMS point is represented by the local minimum of the curve.
}
\label{vcb_3loop_fig}
\end{figure}

If  $b_0=-8.9$, 
the three-loop $b\to c\ell^-\bar\nu_\ell$ inclusive estimated rate 
 is given by
(\ref{rate}) with 
\begin{equation}
S\left[x(\mu),L(\mu)\right]=1-1.67x+\left(-8.9-3.479L\right)x^2+
\left( -50.1-42.4L-7.25 L^2\right)x^3~.
\tlabel{three_loop_S}
\end{equation} 
The coefficients of $x^3$ are those of (\ref{central_coeffs}), as estimated in the previous section.  
Figure \ref{vcb_3loop_fig} displays 
a plot of the $\mu$- (scale-) dependence of the ``reduced'' three-loop rate (\ref{red_rate_def}) with 
$S[x,L]$ given by (\ref{three_loop_S}).  The pole masses $m_b$ and $m_c$ are  
assumed to be $m_b=4.9\,{\rm GeV}$ and $m_c=0.3 m_b=1.47\,{\rm GeV}$ consistent with values used in 
\cite{Czarnecki}.    Figure \ref{vcb_3loop_fig} clearly displays a much flatter $\mu$-dependence than
 the two-loop rate 
plotted in Fig.\ \ref{vcb_2loop_fig}.  In addition to this diminished dependence on the renormalization scale 
$\mu$, the rate plotted in Fig.\ \ref{vcb_3loop_fig} also exhibits a distinct minimum at $\mu=1.0\,{\rm GeV}$.  
At this PMS (minimal-sensitivity) value of $\mu$, the successive terms of the series $S[x,L]$  
exhibit reasonable convergence:
\begin{gather}
S\left[x(1.0\,{\rm GeV}), L(1.0\,{\rm GeV})\right]=1-0.258-0.049+0.020
\tlabel{min_sen_S}\\
\frac{\Gamma_{PMS}}{K}=\frac{\Gamma(1.0\,{\rm GeV})}{K}=1047\,{\rm GeV^5}~.
\tlabel{min_sen_rate}
\end{gather}
Eq.\ (\ref{min_sen_rate}) corresponds to the minimal-sensitivity value for the rate, as discussed in Section 
\ref{two_loop_scale_dep_sec}.

Note that the small three-loop contribution to (\ref{min_sen_S}) can be tuned to zero by making 
only a small change in the choice of the renormalization-scale parameter $\mu$.  
The value of $\mu$ at which the three-loop term vanishes ({\it i.e.,} the value of $\mu$ at which the 
Fig.\ \ref{vcb_2loop_fig} two-loop and Fig.\ \ref{vcb_3loop_fig} three-loop curves intersect) corresponds to the renormalization scale associated with the ``fastest apparent convergence'' (FAC) of the series $S[x,L]$.  
This occurs 
at $\mu=1.18\,{\rm GeV}$:
\begin{gather}
S\left[x(1.18\,{\rm GeV}), L(1.18\,{\rm GeV})\right]=1-0.226-0.058+0
\tlabel{FAC_S}\\
\frac{\Gamma_{FAC}}{K}=\frac{\Gamma(1.18\,{\rm GeV})}{K}=1051\,{\rm GeV^5}~.
\tlabel{Gamma_FAC}
\end{gather}
It is striking that the FAC \cite{Grunberg} and PMS \cite{PMS} criteria predict virtually identical rates; 
moreover, a similar equivalence of rates obtained via these same two criteria is found for the 
estimated three-loop contribution to the 
$b\to u\ell^-\bar\nu_\ell$ rate \cite{V_ub}.  In both  semileptonic processes, 
 the PMS and FAC momentum scales are comparably small.  The PMS and FAC scales ($1.0\,{\rm GeV}$ 
and $1.18\,{\rm GeV}$) for $b\to c\ell^-\bar\nu_\ell$ are respectively 37\% and 44\% of the logarithm 
reference scale $\sqrt{m_bm_c}\cong 2.7\,{\rm GeV}$.  For $b\to u\ell^-\bar\nu_\ell$, $\mu_{_{PMS}}=1.78\,{\rm GeV}$
and  $\mu_{_{FAC}}=1.84\,{\rm GeV}$, numbers which are 42\% and 44\% respectively of  the logarithm reference 
scale $m_b\left(m_b\right)=4.2\,{\rm GeV}$ \cite{vanRit,V_ub}.

\begin{figure}[hbt]
\centering
\includegraphics[scale=0.7]{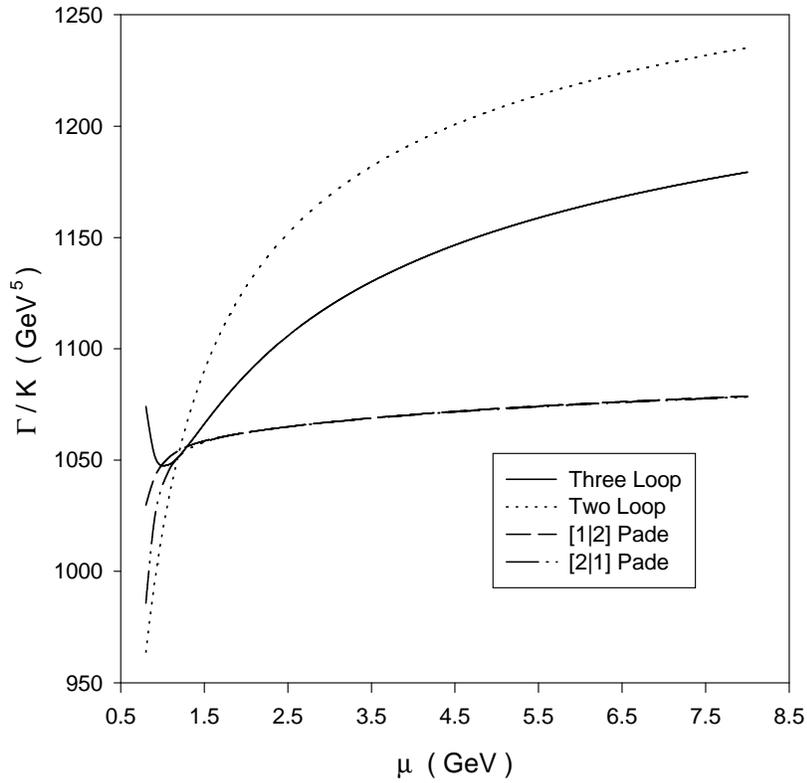}
\caption{Scale dependence of different estimates of the reduced rate in the pole scheme.  The solid curve 
represents  the three-loop estimate presented in Fig.\ \protect\ref{vcb_3loop_fig}, 
and the dotted curve represents the two-loop estimate also presented in Fig.\ \protect\ref{vcb_2loop_fig}.  
The $[1|2]$ 
and $[2|1]$ Pad\'e approximants obtained from the three-loop estimated rate are represented by the dashed curves which 
overlap almost completely above $\mu=1.5\,{\rm GeV}$.  The PMS point is represented by the local minimum of the
three-loop curve, and the FAC point occurs at the intersection of the two- and three-loop curves. 
Note the convergence of {\em all} the estimates near the FAC/PMS points. 
}
\label{vcb_rates_fig}
\end{figure}

Although the estimated three-loop rate plotted in Fig.\ \ref{vcb_3loop_fig} exhibits much less dependence on the
renormalization scale $\mu$ than the two-loop rate of Fig.\ \ref{vcb_2loop_fig}, we anticipate the existence of 
residual $\mu$-dependence as a consequence of the truncation of the series (\ref{three_loop_S}) after three-loop 
order.  A way to eliminate much of this  residual scale dependence is to ``undo the truncation'' 
by choosing an appropriate
Pad\'e approximant to the series (\ref{three_loop_S}).  For example, a $[2|1]$ approximant to the series
(\ref{three_loop_S}) that reproduces its power series to ${\cal O}\left(x^3\right)$ is
\begin{equation}
S^{[2|1]}(x,L)=\frac{1+A_1(L)x+A_2(L)x^2}{1+B_1(L)x}
\tlabel{S_21}
\end{equation}
where
\begin{gather}
A_1(L)=-1.67-\frac{50.1+42.4L+7.25 L^2}{8.9+3.48L}
\tlabel{A1_L}\\
A_2(L)=-(8.9+3.48L)+1.67\,\frac{50.1+42.4L+7.25L^2}{8.9+3.48L}
\tlabel{A2_L}\\
B_1(L)=-\frac{50.1+42.4L+7.25L^2}{8.9+3.48L}
\tlabel{B1_L}
\end{gather}
Similarly, the known series terms (\ref{three_loop_S}) are also reproduced in the power series of the $[1|2]$
approximant 
\begin{gather}
S^{[1|2]}(x,L)=\frac{1+D_1(L)x}{1+E_1(L)x+E_2(L)x^2}
\tlabel{S_12}\\
E_1(L)=-\frac{50.1+42.4L+7.25L^2+1.67\left(8.9+3.48 L\right)}{1.67^2+8.9+3.48L}
\tlabel{E1_L}\\
E_2(L)=\frac{\left(8.9+3.48L\right)^2-1.67\left(50.1+42.4L+7.25L^2\right)}{1.67^2+8.9+3.48L}
\tlabel{E2_L}\\
D_1(L)=E_1(L)-1.67
\tlabel{D1_L}
\end{gather}
In Figure \ref{vcb_rates_fig} we have superimposed plots of the reduced rate (\ref{red_rate_def}) using
\begin{enumerate}
\item 
Eqs.\ (\ref{S_def}) and (\ref{RG_2loop}),  
the two-loop version of $S[x,L]$  leading to the reduced rate also plotted in 
Fig.\ \ref{vcb_2loop_fig},

\item Eq.\  (\ref{three_loop_S}), the three-loop version of  $S[x,L]$ leading to the reduced rate 
plotted in Fig.\ \ref{vcb_3loop_fig}, 

\item $S^{[2|1]}$, as determined via (\ref{S_21}--\ref{B1_L}), and

\item  $S^{[1|2]}$, as determined via (\ref{S_12}--\ref{D1_L}).

\end{enumerate}
The vertical scale of Figure \ref{vcb_rates_fig} is magnified compared to that of Figs.\ \ref{vcb_2loop_fig} and
\ref{vcb_3loop_fig} in order to accentuate the differences between reduced rates 
obtained for each of the above scenarios.
 We observe from Fig.\  \ref{vcb_rates_fig}
 that both Pad\'e-approximant versions of the rate coincide after    $\mu  = 1.5\,{\rm GeV}$ and are considerably
flatter than the rate devolving from the  three-loop version of $S[x,L]$.  Indeed, the three-loop
reduced rate is itself quite stable, increasing slowly from $1066\, {\rm GeV^5}$   to $1180\,{\rm  GeV^5}$
   as  $\mu$     increases
from $1.5\,{\rm GeV}$ to $8\, {\rm GeV}$. Nevertheless, the two Pad\'e approximant versions of $S[x,L]$ 
vary only
minimally over the same range of  $\mu$, increasing from $1058\,{\rm GeV^5}$    at   $\mu  = 1.5\, {\rm GeV}$
 to $1079\, {\rm GeV^5}$  
at   $\mu = 8\, {\rm GeV}$. Thus Pad\'e-improvement of the three-loop rate virtually eliminates 
the residual scale
dependence of the naively truncated expression. Such use of Pad\'e approximants to eliminate residual scale 
dependence is also evident in  Fig. 3 of ref.\ \cite{V_ub} for the 
$b\to u\ell^-\bar\nu_\ell$ rate, and has been previously discussed in the context of the Bjorken sum-rule
\cite{Bjorken} as well as in more general terms \cite{Gardi}.   
 Of particular interest, however, is the convergence of all four curves in Fig.\ \ref{vcb_rates_fig} to virtually the same PMS/FAC point.  This convergence lends further support to the PMS/FAC estimates (\ref{min_sen_rate}),
(\ref{Gamma_FAC}) for the reduced rate. 

\section{Discussion}\label{discuss_sec}
  The entire analysis presented in Section \ref{three_loop_scale_dep_sec} can be repeated using the extreme values 
$b_0=-8.6$ and
$b_0=-9.2$, as estimated in \cite{Czarnecki} (again, see footnote \ref{czarnecki_note}), utilising (\ref{high_coeffs}) and (\ref{low_coeffs}) 
for the appropriate determinations
of three-loop coefficients in conjunction with the known  values of $a_0$ and $b_1$   (\ref{RG_2loop}).  One
finds the uncertainty in $b_0$   is reflected in a   $\pm 14\, {\rm GeV^5}$ spread in the 
PMS/FAC value $1050\, {\rm GeV^5}$ for
the reduced rate.

  Other sources of theoretical uncertainty arise from    $\alpha_s\left(m_\tau\right)= 0.33 \pm 0.02$ 
\cite{aleph,steele}, 
$m_b^{pole}= (4.9\pm   0.1)\, {\rm GeV}$ \cite{hoang}, and the error that may occur in estimating $c_0$. 
We estimate that Pad\'e determinations of $c_0$   are subject to errors comparable to those of 
Pad\'e
determinations of the RG-accessible coefficients $c_1$   and $c_2$; {\it e.g.,} 
$\left|\left(c_1^{Pade}-c_1^{RG}\right)/c_1^{RG}\right|\cong 7\%$. 
If we are conservative and estimate the uncertainty of $c_0$   to be double that of $c_1$,
\begin{equation}
\left|\frac{\delta c_0}{c_0}\right|\cong 14\%~,
\tlabel{c0_err}
\end{equation}
the corresponding uncertainty in the reduced rate is $\pm 38\, {\rm GeV^5}$. 
Consequently, our estimate of the
purely perturbative three-loop order $b\to c\ell^-\bar\nu_\ell$  rate, as defined by (\ref{red_rate_def}), 
is 
\begin{equation}
\frac{\Gamma^{pert}}{K}=\left(1050\pm 14\pm 44\pm 115\pm 38\right)\,{\rm GeV^5}~,
\tlabel{Gamma_err}
\end{equation}
where the listed theoretical uncertainties respectively devolve from the uncertainty in $b_0$,  
$\alpha_s\left(m_\tau\right)$, 
$m_b^{pole}$, and $c_0$.

  Nonperturbative (NP) contributions to the rate may be extracted from Eq.\ (5.8) of 
\cite{bagan}, and correspond to  the following additional contributions to the series $S[x,L]$:
\begin{equation}
\Delta S^{NP}=\frac{\lambda_1+3\lambda_2}{2m_b^2}-
\frac{6\left(1-\frac{m_c^2}{m_b^2}\right)^4\lambda_2}{m_b^2\,F\left(\frac{m_c^2}{m_b^2}\right)}~,
\tlabel{S_NP}
\end{equation}
where the form factor $F$ is given by (\ref{form_fac}) and where
\begin{equation}
-0.5\,{\rm GeV^2}\le\lambda_1\le 0~,~\lambda_2=0.12\,{\rm GeV^2}~.
\tlabel{vevs}
\end{equation}
This additional NP contribution entails a  $\sim 6\%$ reduction in the reduced rate (\ref{Gamma_err}):
\begin{equation}
\frac{\Gamma}{K}=\frac{\Gamma^{pert}}{K}-\left(58.5\pm 5.5\right)\,{\rm GeV^5}=
\left(992\pm 217\right)\,{\rm GeV^5}~,
\tlabel{final_Gamma}
\end{equation}
where the independent sources of uncertainty in (\ref{Gamma_err}) and in the intermediate step of 
(\ref{final_Gamma}) have been
combined additively.

If we identify the predicted $b\to c\ell^-\bar\nu_\ell$ decay rate with the inclusive semileptonic process
$B\to X_c\ell^-\bar\nu_\ell$  [$\ell=e$ or $\mu$, but not their sum], we can then relate the aggregate
theoretical uncertainty in (\ref{final_Gamma}) to the concomitant theoretical uncertainty in the
determination of $\left|V_{cb}\right|$:
\begin{equation}
\left|V_{cb}  \right|=\left[\frac{192\pi^3\hbar\, 
{\rm BR}\left(B\to X_c\ell^-\bar\nu_\ell\right)}{G_F^2\tau_{_B}
\left[\left(992\pm 217\right)\,{\rm GeV^5}\right]}\right]^{1/2}
\tlabel{Vcb_expr}
\end{equation}
To factorize experimental and theoretical uncertainties, we employ recent central values for the average $B$ lifetime $\tau_{_B}$ \cite{Battaglia} and the $B\to X_c e^-\bar\nu_e$ branching ratio \cite{BELLE} to 
rewrite (\ref{Vcb_expr}) in the following form:
\begin{equation}
\left|V_{cb}  \right|=\left(0.0453^{+0.0060}_{-0.0043}\right)\,
\left(\frac{1.564\times 10^{-12}\,{\rm s}}{\tau_{_B}}\right)^{1/2}\,
\left(\frac{{\rm BR}\left(B\to X_c\ell^-\bar\nu_\ell\right)}{0.1105} \right)^{1/2} 
\tlabel{Vcb_final}
\end{equation}
The first factor in (\ref{Vcb_final}) reflects the summed theoretical uncertainties 
in (\ref{final_Gamma}), 
which are separately broken down in (\ref{Gamma_err}).  We have been conservative in identifying and assessing the magnitude of each such independent source of error--- the theoretical uncertainty estimated from a
(truncated) two-loop calculation of the $b\to c\ell^-\bar\nu_\ell$ rate should be {\em larger} than that obtained by us in (\ref{Vcb_final}). 

    Note also that we can compare our   
$\Gamma^{pert}/K=1050\, {\rm GeV^5}$   central value
estimate (\ref{Gamma_err}) for the reduced rate (exclusive of NP effects) with the corresponding estimate one would obtain using low-scale
masses \cite{Czarnecki}:\footnote{The ${\cal O}\left(\alpha_s^2\right)$ 
coefficient  in 
(\protect\ref{low_scale_rate}) [A. Czarnecki, personal communication] differs slightly  
from the value $-2.65\pm 0.4$ appearing in ref.\ \protect\cite{Czarnecki}.
  }
\begin{equation}
\frac{\Gamma^{pert}}{K}=\tilde m_b^5\,F\left(\frac{\tilde m_c^2}{\tilde m_b^2}\right)
\left[1-1.14\frac{\alpha_s\left(\sqrt{\tilde m_b\tilde m_c}\right)}{\pi}
-\left(3.5\pm 0.3\right)\left(\frac{\alpha_s\left(\sqrt{\tilde m_b\tilde m_c}\right)}{\pi} \right)^2
\right]~.
\tlabel{low_scale_rate}
\end{equation}
If we utilise the low-scale mass values $\tilde m_b= 4.64\,{\rm GeV}$, $\tilde m_c= 1.25\, {\rm GeV}$, 
as quoted in \cite{Czarnecki} from ref.\ \cite{Bigi} and find via devolution from 
$\alpha_s\left(m_\tau\right)=0.33$
 that   $\alpha_s\left(\sqrt{\tilde m_b\tilde m_c}\right)/\pi= 0.091$, we observe that the
rate predicted via (\ref{low_scale_rate}) is $1097\, {\rm GeV^5}$, 
an answer in surprisingly good agreement with our $1050\,{\rm GeV^5}$ three-loop estimate in the pole-mass renormalization scheme.

We conclude by noting that the RG equation (\ref{RG_inv}, \ref{RG_eqn}) may be used to determine 
additional higher-order corrections to the decay rate (\ref{rate}).  The leading-log corrections to all
orders in $x$ are determined  by the one-loop $\beta$-function; next-to-leading-log corrections to all
orders in $x$ are determined by the two-loop $\beta$-function {\it etc}.  Although we have made use of 
RG-invariance to ${\cal O}\left(x^3\right)$ in present work, it is in fact possible to incorporate these logarithmic corrections to all subsequent orders.  The full exploitation of RG-invariance within 
perturbative-QCD 
expressions   for physical processes is presently under study.

\smallskip
\noindent
{\bf Acknowledgments:} We are grateful for the hospitality of the KEK Theory Group, where this research 
was conducted with financial  support
from 
the  International Opportunity Fund of the
 Natural Sciences and Engineering Research Council of Canada (NSERC) 
and the International Collaboration Programme
2001 of Enterprise Ireland.

\clearpage

\clearpage

\clearpage

\clearpage

\clearpage

\end{document}